\documentclass[fleqn,usenatbib]{mnras}

\usepackage{newtxtext,newtxmath}

\usepackage[T1]{fontenc}
\usepackage{ae,aecompl}

\usepackage{graphicx}	
\usepackage{amsmath}	
\usepackage{amssymb}	
\usepackage{xcolor}

\title[Disrupted Planetary Discs Inside the Core of an AGB Star] 
{Hydrodynamic Simulations of Disrupted Planetary Accretion Discs Inside the Core of an AGB Star}
\author[Guidarelli et al.]{G. Guidarelli$^{1}$\thanks{E-mail: gcg3642@g.rit.edu },
J. Nordhaus$^{1,2}$\thanks{E-mail: nordhaus@astro.rit.edu},
L. Chamandy $^{3}$,
Z. Chen$^{4}$,
E. G. Blackman$^{3}$,
\newauthor A. Frank$^{3}$,
J. Carroll-Nellenback$^{3,5,6}$,
B. Liu$^{3,5,6}$,
\\
$^{1}$Center for Computational Relativity and Gravitation, Rochester Institute of Technology, Rochester, NY, 14623, USA\\
$^{2}$National Technical Institute for the Deaf, Rochester Institute of Technology, Rochester, NY, 14623, USA\\
$^{3}$Department of Physics and Astronomy, University of Rochester, Rochester, NY  14627, USA\\
$^{4}$Department of Physics, University of Alberta, Edmonton, AB, T6G 2E1, Canada \\
$^{5}$Center for Integrated Research Computing, University of Rochester, Rochester, NY  14627, USA\\
$^{6}$Laboratory for Laser Energetics, University of Rochester, Rochester, NY  14623, USA
}

\date{Accepted XXX. Received YYY; in original form ZZZ}

\pubyear{2019}

\begin{document}
\label{firstpage}
\pagerange{\pageref{firstpage}--\pageref{lastpage}}
\maketitle

\begin{abstract}
Volume complete sky surveys provide evidence for a binary origin for the formation of isolated white dwarfs with magnetic fields in excess of a MegaGauss. Interestingly, not a single high-field magnetic white dwarf has been found in a detached system suggesting that if the progenitors are indeed binaries, the companion must be removed or merge during formation. An origin scenario consistent with observations involves the engulfment, inspiral, and subsequent tidal disruption of a low-mass companion in the interior of a giant star during a common envelope phase.  Material from the shredded companion forms a cold accretion disc embedded in the hot ambient around the proto-white dwarf.  Entrainment of hot material may evaporate the disc before it can sufficiently amplify the magnetic field, which typically requires at least a few orbits of the disc.
Using three-dimensional hydrodynamic simulations of accretion discs with masses between 1 and 10 times the mass of Jupiter inside the core of an Asymptotic Giant Branch star, we find that the discs survive for at least $10$ orbits (and likely for 100 orbits), sufficient for strong magnetic fields to develop.
\end{abstract}




\section{Introduction}

White dwarfs (WDs) are the terminal end products of stellar evolution for main-sequence stars with initial masses $\lesssim$8 $M_\odot$. Zeeman spectroscopy has revealed a peculiar subset of WDs with abnormally high surface-averaged magnetic fields; aptly termed high-field magnetic white dwarfs (HFMWD) \citep{Preston_1970}.
Constituting $\sim$10$\%$ of all WDs, HFMWDs have magnetic fields of $\sim$$\mathrm{10^4 - 10^9\, Gauss}$ while typical WDs have a magnetic field of $\sim$$\mathrm{10^1 - 10^3\, Gauss}$ \citep{Schmidt_2003,Liebert_2003,Kepler_2013,Kepler_2018,Ferrario2015}. They are more massive than generic WDs \citep{Kawka_2007} and none have been found in detached binary systems with M dwarf companions \citep{Liebert_2005, Silvestri_2006,Silvestri_2007}.

The origin of the strong magnetic fields present in HFMWDs is debated. There are arguments the magnetic field is a remnant from the progenitor main-sequence star \citep{Angel_1981,Wickram_2005} and more recently the plausibility of magnetic field generation via an internal dynamo as the white dwarf crystallizes has been explored \citep{Isern_2017}. However, neither of these theories easily explain the mass disparity or the lack of HFMWDs in non-interacting binaries. An alternative is that the magnetic field is correlated with close binary interactions such as mergers or common envelope evolution (CEE).

CEE creates a pathway for reducing binary separations and leads to some of the most interesting and energetic events in the observable universe \citep{Paczynski1976,Nordhaus2006,Staff_2016,Belczynski:2016sf,wilson2019}. CEE may be a key process in forming a HFMWD as it can lead to WD-WD mergers and the engulfment and accretion of lower-mass companions. A hot corona formed after a WD-WD merger might last long enough to generate a strong magnetic field consistent with observations of HFMWDs in the solar neighborhood \citep{Garc_a_Berro_2012}. This scenario requires two common envelope interactions as both main-sequence stars evolve to become white dwarfs.  The consistency of the mass distribution of the merged WD remnants with the mass distribution of isolated HFMWDs remains a subject of debate.  

Alternatively, the magnetic field might also be the result of CEE between the pre-WD AGB star and a low-mass companion \citep{Nordhaus:2011aa}. As the AGB star expands to form the CE, the relatively slow moving circumbinary material would dynamically drag on the companion thereby reducing its orbit. If the companion mass is sufficiently low, the liberated orbital energy during inspiral will not exceed the envelope's binding energy, resulting in continued orbital decay until the companion reaches the tidal shredding radius. At this radius, the self gravity of the companion is overwhelmed by the differential potential across its surface resulting in extreme deformation and eventual disruption. The subsequent accretion disc formed by the tidally shredded material could amplify (via a dynamo), advect and anchor a magnetic field to the surface of the white dwarf core.  A long Ohmic decay timescale ensures that a HFMWD emerges at a later stage of evolution after the star has shed its outer layers.  We focus on this scenario, as it is consistent with the stringent observational constraints provided by the Sloan Digital Sky Survey \citep{Nordhaus:2011aa}.

Analytic estimates for the magnetic fields generated via a dynamo operating in discs formed from $\sim$1-500 $M_{\rm J}$ companions\footnote{$M_{\rm J}$ is the mass of Jupiter.} were sufficient to explain the full range of observed HFMWDs \citep{Nordhaus:2011aa}.  However,  this scenario  requires the disc to survive at least several rotation times for the fields to amplify.  Generation of strong magnetic fields is a necessary but not necessarily sufficient condition to produce a HFMWD.  The disc must also persist and sustain steady-state magnetic fields long enough to accrete onto the proto-WD. In this purely hydrodynamic study, we focus on stability of the disc in the presence of the harsh interior of the AGB star and leave details of magnetic field amplification and advection to future work.  

The disc is initially cold ($\sim$$10^{4-5}$ K) and dense ($\sim$$1$ ${\rm g}$ ${\rm cm^{-3}}$), compared to the hot ($\sim$$10^{7-8}$ K) and less dense ($\sim$$10^{-2}$ ${\rm g}$ ${\rm cm^{-3}}$) interior of the AGB star.  Vertical shear and steep temperature gradients may threaten disc stability as entrainment of hot  material will eventually fully mix the disc into the stellar envelope. The central question that motivates this work is to determine if the discs survive the AGB interior long enough so that a strong magnetic field could develop. 

In this paper, we present hydrodynamic simulations of accretion discs formed from the disruption of planetary companions in the interior of a $2 M_\odot$ AGB star.  Section 2 details some additional aspects of the formation scenario while Section 3 outlines our numerical approach, setup and simulation parameters. Section 4 presents our simulation results while we conclude and discuss future directions in Section 5.

\section{Additional Aspects of the Formation Scenario}

Before detailing our numerical approach, we note that this scenario requires the companion to survive until the point at which it is tidally disrupted and forms a disc.  While it is likely that low-mass companions survive inspiral until they are tidally disrupted, there are a few effects that might lead to the planet's destruction before this point which we comment on here.  

\cite{Villaver_2007}  previously suggested that thermal evaporation during a common envelope phase destroys companions of less than 15-Jupiter masses.  However, this scenario requires that the radiative flux penetrate the large optical depth of the planet on a timescale shorter than orbital decay.  At $10^{11}$ cm from the center of an extreme-AGB star, the evaporation timescale is on the order of $\sim$$10^{3-4}$ years compared to an orbital decay timescale of a few years, suggesting that the planet likely shreds before significant evaporation occurs \citep{Watson_1981,Murray-Clay_2009, Lopez_2012}.  Furthermore, brown dwarfs around white dwarfs in post-CE orbits show no evidence that the common envelope phase affected their masses \citep{Maxted_2006,Casewell_2018,Longstaff_2019}.  In contrast to brown dwarfs, all but the most massive planets are unlikely to survive the common envelope phase based on theoretical arguments and observational searches for planets in white dwarf habitable zones \citep{Nordhaus:2013aa,Sandhaus_2016,Xu_2015,Van_Sluijs_2017,Vanderburg_2015}.

Another possibility is that the orbital energy dissipated by drag might be communicated to the interior of the planet,  unbinding it before it tidally disrupts.  This is unlikely to occur via mechanical means in light of the large density contrast between the AGB envelope and the planetary interior.  Aerodynamic drag exerts a steady pressure on the leading face of the planet, and turbulent mixing will be inhibited by the steep entropy gradient at the planet's surface.  While it appears likely that even low-mass planets inspiral until they are tidally disrupted, high-resolution CE simulations with self-gravitating planets in hydrostatic equilibrium (HSE) are needed to ultimately verify these assumptions.

\begin{figure}
    \centering
    \includegraphics[scale=0.185]{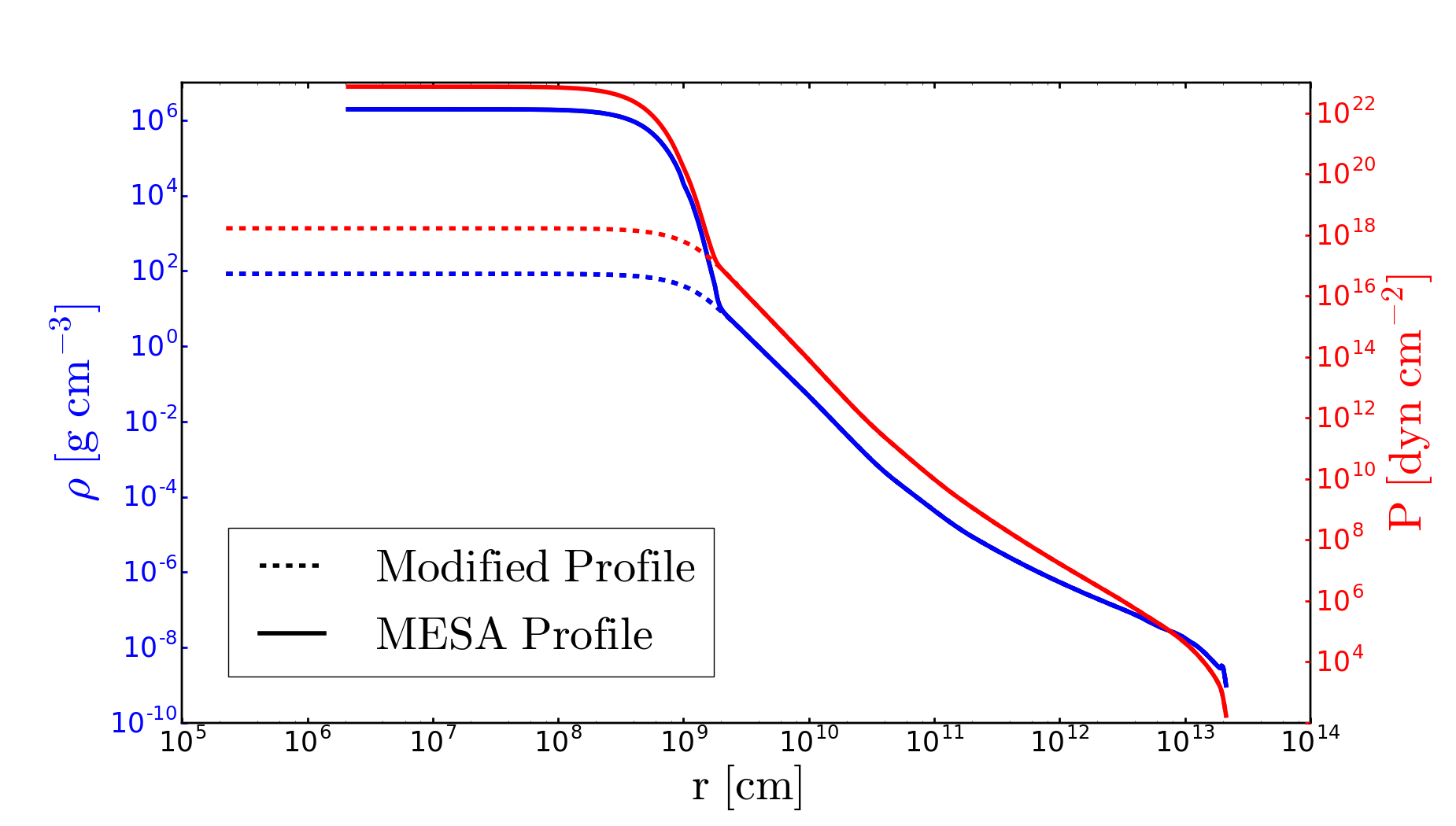}
    \caption{Stellar density and pressure profiles before and after modification.} 
    \label{fig:stellar_interior}
\end{figure}
Additionally, we note that the disc masses considered in this work are less than their typical progenitor companions.  According to StarSmasher simulations of tidal interactions of Jupiter-mass companions around low-mass/intermediate-mass black holes, $\sim$$50\%$ of the companion mass remains bound after disruption \citep{Perets_2016}. We expect a similar retention ratio for our systems. 

The fraction of MS stars with one or more companions has been estimated to be $\sim$$0.44$ \citep{Raghavan_2010}. The mass fraction distribution of intermediate-mass main-sequence stars follows $f(q)$ $\mathrm{\propto}$ $q^{\sigma}$ where $q$ is the mass ratio and $\sigma = -0.5\pm0.2$ \citep{Dushene_2013}. This implies that $\sim$$25\%$ of the main sequence stars with companions have $q\gtrsim0.125$. This is on the order of mass ratios predicted to be involved with this common envelope scenario. Therefore, the fraction of main-sequence stars with low-mass companions is of order the fraction of HFMWDs.

\section{Numerical Methods}
We performed three-dimensional hydrodynamic simulations of accretion discs ($\mathrm{1-10} {\rm M_J}$) around the core of a $\mathrm{2 M_\odot}$ AGB star. The discs were given an initial Keplerian rotation profile with a radius of $\mathrm{2\times 10^{10}\,cm}$ and height of $\mathrm{5 \times 10^9\,cm}$.  The radial extent of the accretion disc is chosen to be the tidal shredding radius, which is the location where the differential gravitational potential of the proto-WD core exceeds the self gravity of a constant density sphere, and is given by:
\begin{equation}
    R_\mathrm{s} \simeq R_\mathrm{c} \left(\frac{2 M}{M_\mathrm{c}}\right)^{1/3}
\end{equation}
where $R_\mathrm{c}$ and $M_\mathrm{c}$ are the radius and  mass of the companion, $M$ is the mass of the primary interior to $R_\mathrm{s}$ \citep{Nordhaus2006}. 

The initial disc masses were selected according to previous analytic estimates of systems that produce HFMWDs \cite{Nordhaus:2011aa}. The ambient AGB density distribution was determined with the one-dimensional stellar evolution code MESA. With MESA, we evolved a $2 M_\odot$ zero-age-main-sequence star with solar metalicity ($z=0.02$) through all phases of its stellar lifetime until it began cooling as a white dwarf. The density and pressure profiles were extracted at the time when the star's radius was at its maximum.  This is a reasonable time at which a companion would be engulfed as the physical volume of the star is maximized and tidal torques are the strongest \citep{Nordhaus:2010aa,Nordhaus:2013aa}.

\begin{figure*}
    \centering
    \includegraphics[height=5cm]{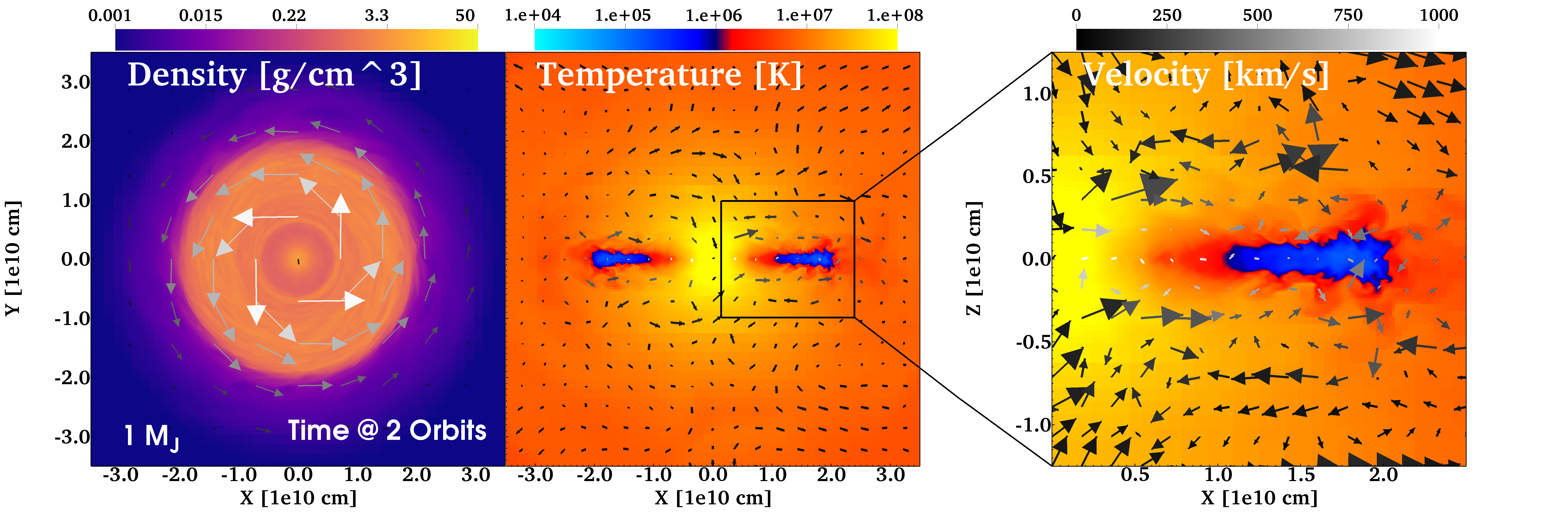}
    \includegraphics[height=5cm]{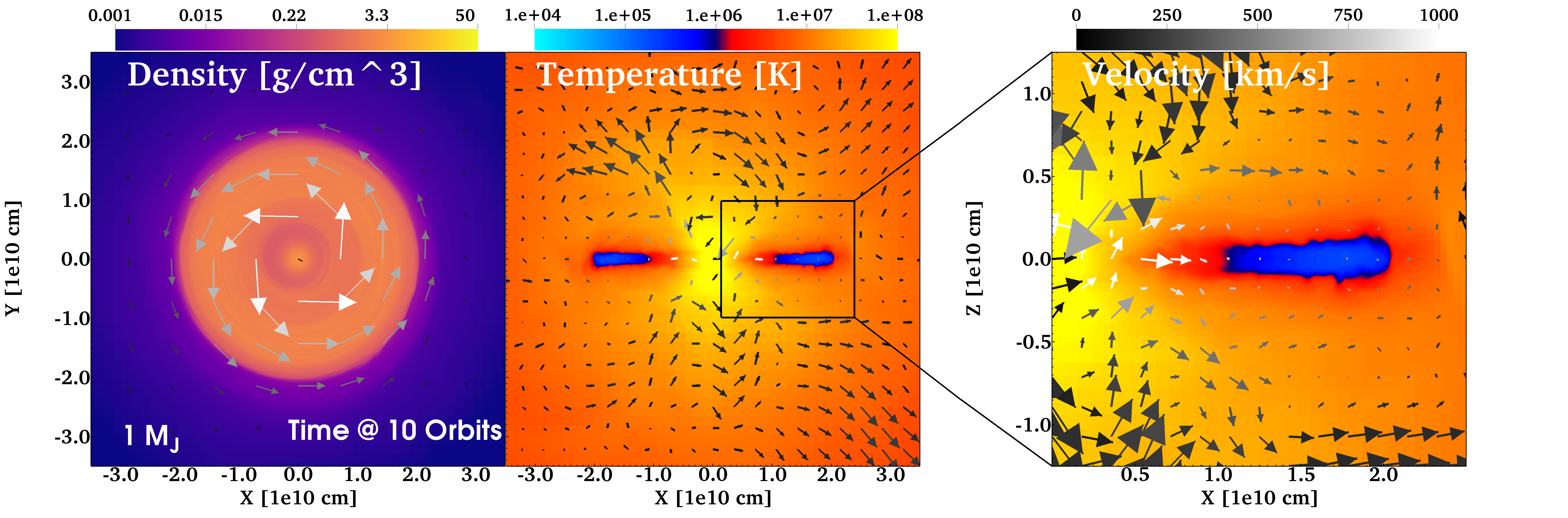}
    \caption{Evolution of the one Jupiter mass disc at 2 and 10 orbits. The left-most panel is a mid-plane slice of the density of the disc from a top down view. The other panels show a slice of the temperature of the disc from the side with the right-most panel a magnified view indicated by the black box. Velocity vectors are shown on all plots with magnitude indicated by colour and the projected magnitude by the length.}
    \label{fig:1MJ}
\end{figure*}

In order to ensure that boundary conditions do not affect the disc dynamics it is necessary to resolve beyond the disc by at least an order of magnitude ($\sim\mathrm{10^{11}\, cm}$). The simulation box size combined with limited computation power means that the innermost region of the AGB density profile cannot be properly resolved. We employ a point particle at the center of the grid to accurately match the gravitational potential. This requires a modification to the original profile to conserve total mass and maintain hydro-static equilibrium. We use the modified Lane-Emden equation to adjust the profile:
\begin{equation}
    \frac{1}{\xi} \frac{d}{d\xi} \left(\xi^2 \frac{d\theta}{d\xi}+\xi^2 \frac{g_c(\alpha \xi)}{4 \pi G \rho_c \alpha}\right) + \theta^n = 0.
\label{laneemden}
\end{equation}

The Lane-Emden equation is modified by the second term on the left-hand side in Equation~\ref{laneemden}, representing the gravitational attraction of the point particle \citep{Ohlmann2017}.  The polytropic index is given as $n$, with the re-scaled radial coordinate, $\mathrm{\xi}$, and density, $\mathrm{\theta}$, defined by:
\begin{equation}
     \xi \equiv \frac{r}{\alpha}
\end{equation}
 and
\begin{equation}
     \theta \equiv \left(\frac{\rho}{\rho_{c}}\right)^{2/3}
\end{equation}
where
\begin{equation}
     \alpha^2 \equiv \frac{5\rho_c^{-1/3}}{8\pi G}.
\end{equation}
Lastly, $\mathrm{g_c}$ is the smoothed gravitational acceleration from the point particle and $y\equiv r/h$:
\begin{align} \label{gc}
g_c(r) = 	G m_c \left\{ \begin{array}{cc} 
                \frac{1}{r^2} & \hspace{3mm} r>=h \\
                \frac{y(\frac{64}{3}+y(-48+y(\frac{192}{5}+y\frac{-32}{3})))-\frac{2}{30y^2}}{h^2} & \hspace{3mm} h/2<=r<h \\
               \frac{y(\frac{32}{3}+y^2(\frac{-192}{5}+32y))}{h^2} & \hspace{3mm} r<h/2 \\
                \end{array} \right.
\end{align}

Equation~\ref{laneemden} was solved using a third order Runge-Kutta integrator such that solutions were accepted when the slope of the density profile matched at the smoothing radius. The resultant profiles in Figure~\ref{fig:stellar_interior} have a reduced constant central density where the profile was previously underresolved. The added point particle ensures the total mass inside the smoothing radius remains the same as the initial model.  

A point particle also requires the choice of sub-grid model for losing and gaining mass. The current sub-grid models for accretion available in {\sc AstroBEAR} remove thermal energy from the grid, a fact that is only appropriate in contexts where the gas exterior to the point particle is optically thin.  In our environment this would lead to unphysical accretion rates \citep{Krumholz_2004,Federrath_2010,Chamandy:2018aa}. Therefore, we choose a non-accreting point particle which allows the gas to accumulate naturally in the center. As the pressure builds, the accretion rate decreases until it is eventually halted.  Advection of the field and formation of a HFMWD requires that the central pressure be removed, most likely through a bipolar outflow or jet \citep{Nordhaus:2011aa} and that the quasi-steady state field sustain in the disc until that time.

The fluid in our simulation abides by
\begin{equation}
    \mathrm{\frac{\partial \rho}{\partial t} + \nabla \cdot (\rho \mathbf{v}) = 0},
\end{equation}
\begin{equation}
     \frac{\partial \rho \mathbf{v}}{\partial t} + \nabla \cdot (\rho \mathbf{v v}) = -\nabla P - \rho g_c(r)\hat{r},
\end{equation}
and
\begin{equation}
     P = \tilde{n} k_\mathrm{b} T,
\end{equation}
where $\rho$ is the density, $t$ is time, $\mathbf{v}$ is the velocity, $p$ is the pressure, $r$ is the radial coordinate, $\tilde{n}$ is the number density of particles, $k_\mathrm{b}$ is Boltzmann constant and $T$ is the temperature. With monatomic the $\gamma$-law gas equation of state where $\gamma = 5/3$.

The total simulated time for each disc mass is $\sim 22000\mathrm{\,s}$, approximately 10 orbits. The simulations were completed with the three-dimensional, multi-physics, AMR code, {\sc AstroBEAR}. {\sc AstroBEAR} utilizes a Riemann solver to solve the fluid equations.

We use $128^3$ computational cells with $5$ levels of refinement for an effective resolution of $4096^3$ with extrapolated boundary conditions. The simulated box side lengths are $2\times 10^{11}$ cm which means the smallest distance resolved is $\sim$$10^8$ cm. This resolution accurately resolves the modified ambient profile transition which is important for maintaining HSE and is typically two orders of magnitude smaller than the smallest grid cells used in CE simulations that resolve the full primary star \citep{Ohlmann2015,Chamandy:2018aa,Chamandy_2019,Passy2012}. The computations were run on Stampede 2 at TACC \citep{XSEDE}.

\section{Results}

\begin{figure*}
    \centering
    \includegraphics[height=5cm]{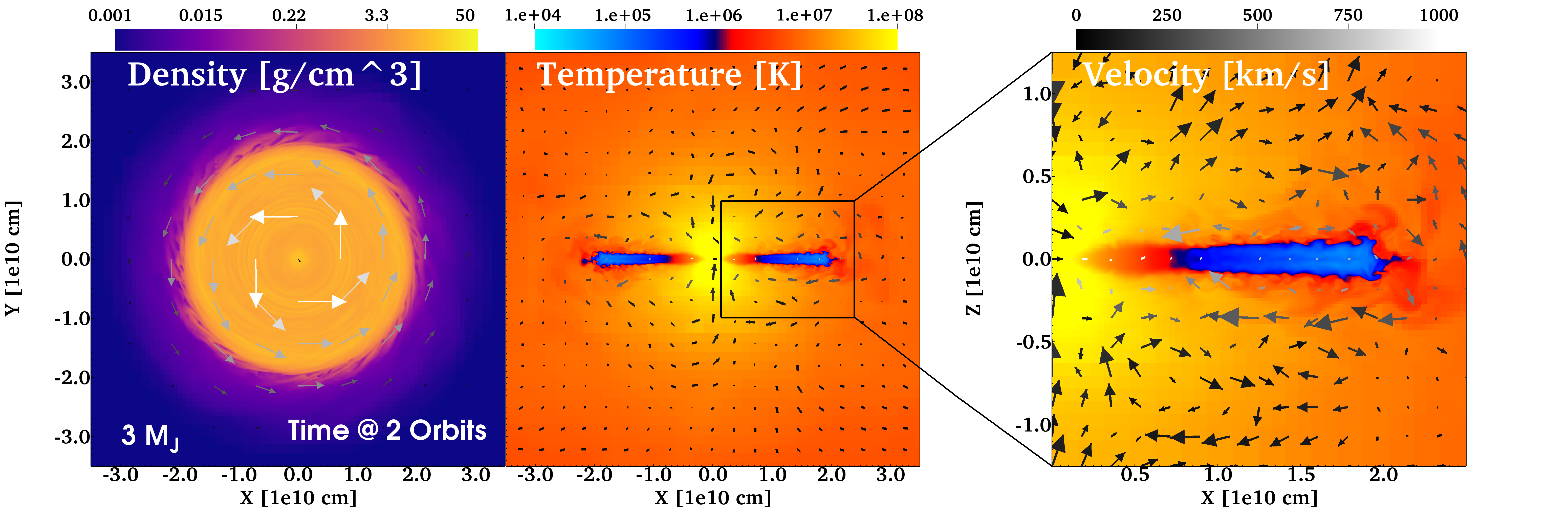}
    \includegraphics[height=5cm]{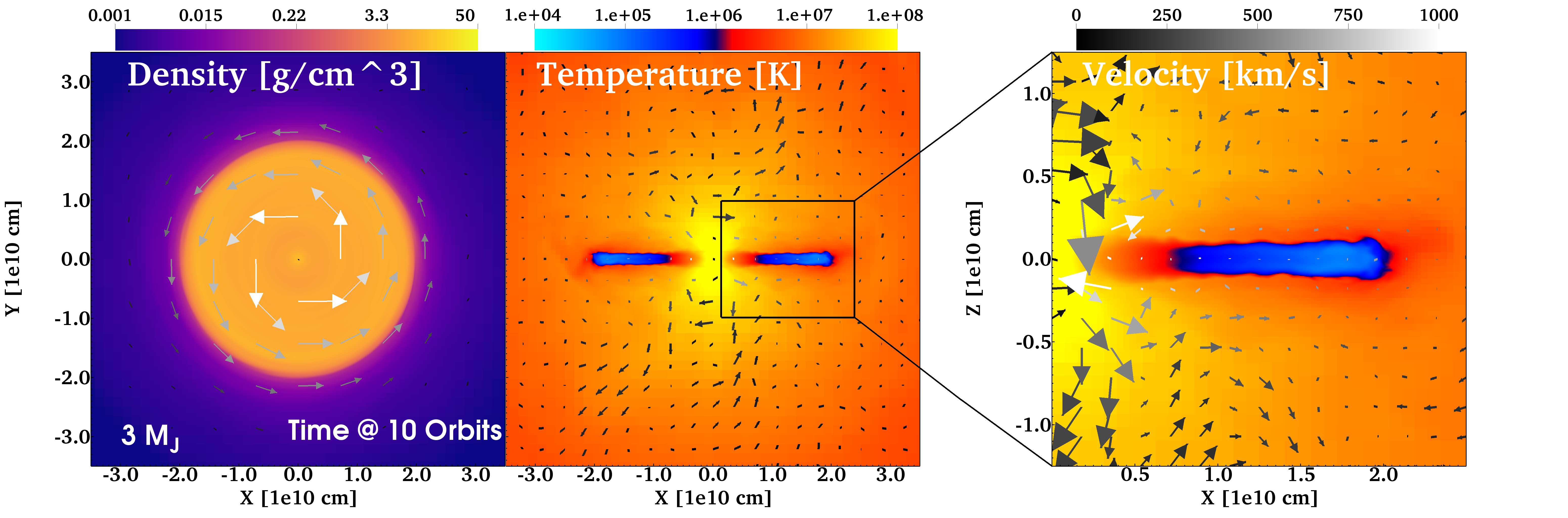}
    \caption{Evolution of the three Jupiter mass disc at 2 and 10 orbits. (Similar to \ref{fig:1MJ})
    \label{fig:3MJ}}
\end{figure*}

\begin{figure*}
    \centering
    \includegraphics[height=5cm]{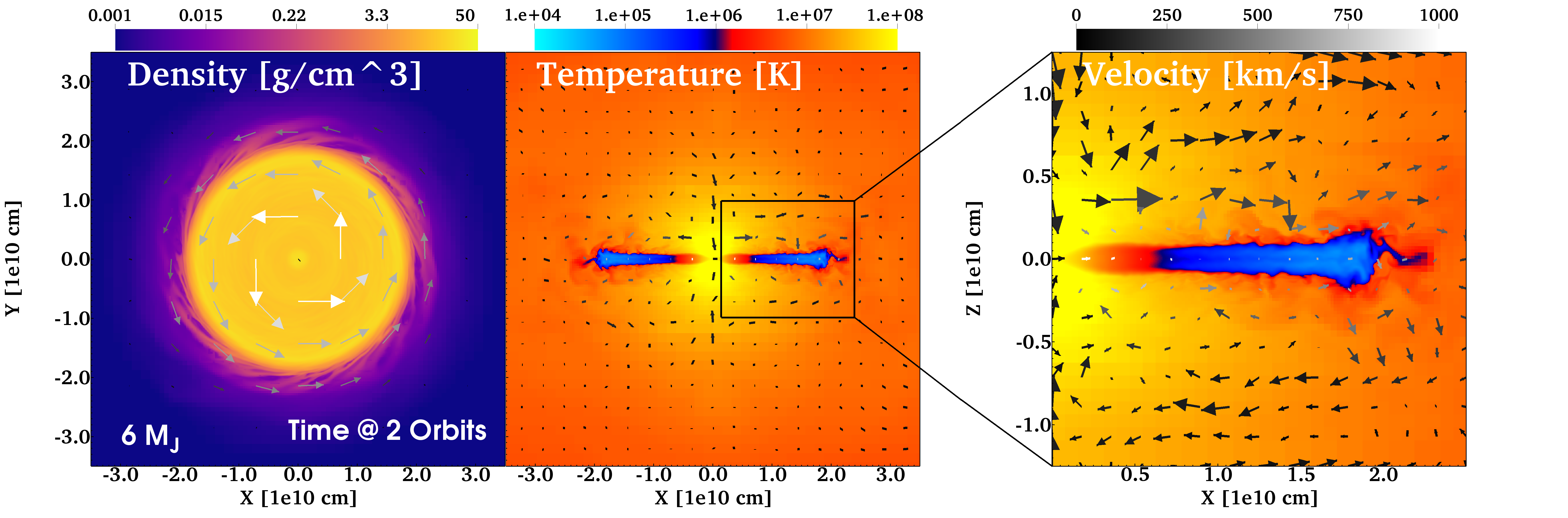}
    \includegraphics[height=5cm]{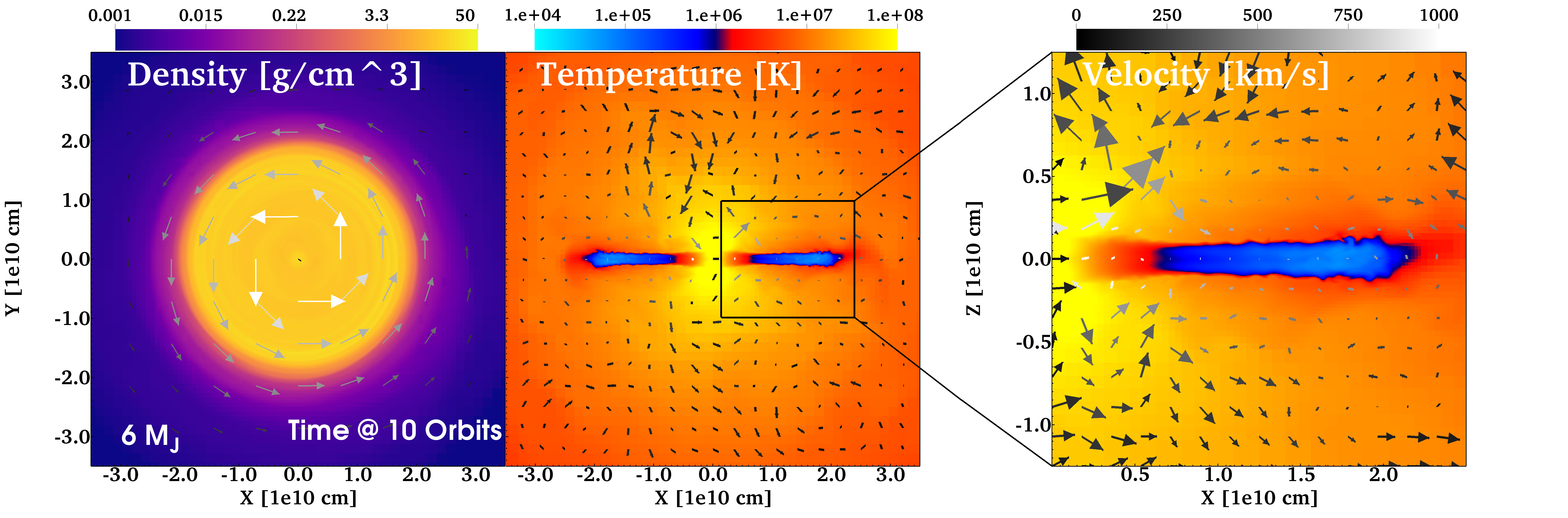}
    \caption{Evolution of the six Jupiter mass disc at 2 and 10 orbits. (Similar to \ref{fig:1MJ})
    \label{fig:6MJ}}
\end{figure*}

\begin{figure*}
    \centering
    \includegraphics[height=5cm]{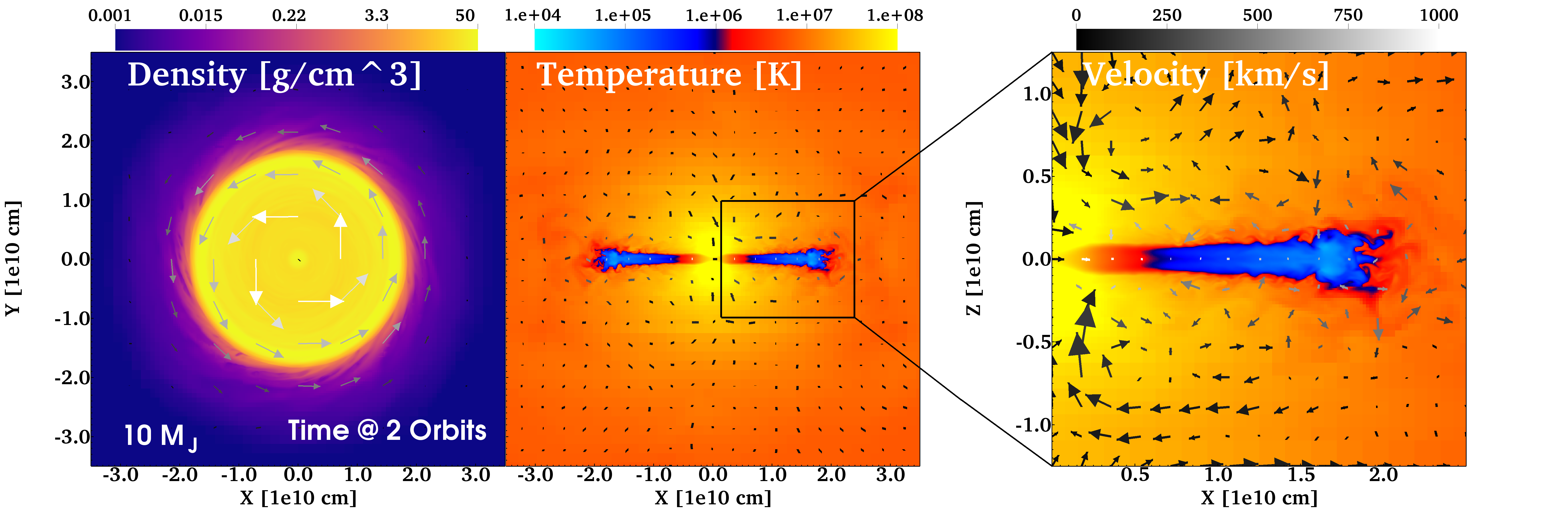}
    \includegraphics[height=5cm]{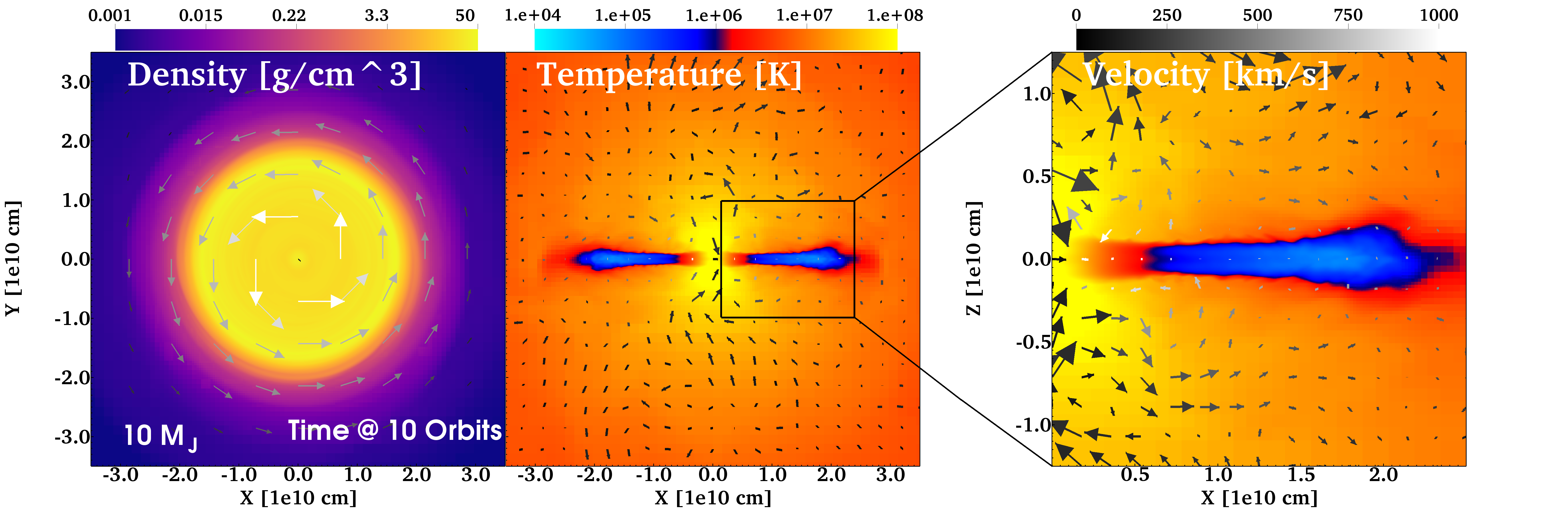}
    \caption{Evolution of the ten Jupiter mass disc at 2 and 10 orbits. (Similar to \ref{fig:1MJ})
    \label{fig:10MJ}}
\end{figure*}

Figures \ref{fig:1MJ}-\ref{fig:10MJ} present density and temperature snapshots from each simulation at 2 and 10 orbits\footnote{An orbit is defined as the rotation period of the disc's outer radius at the start of the simulation.}. In each figure, the left panel shows a face-on view of the density of the mid-plane of the disc.  The middle panel presents an edge-on view of temperature while the right panel details the zoomed-in region of the middle panel.  The top three panels show the disc structure with corresponding velocity vectors at 2 orbits while the bottom three panels present the same information at 10 orbits.

The temperature plots for each of the discs show pronounced Kelvin-Helmholtz instabilities at 2 orbits.  Comparing the profile of each disc at 2 and 10 orbits, it is clear that the discs expand vertically-upward and radially-outward with the more massive discs expanding the most. The ambient velocities initially appear stochastic at 2 orbits and develop more structure at 10 orbits as the gas is spun up and circulates with the disc.

\begin{figure}
    \centering
    \includegraphics[scale=0.17]{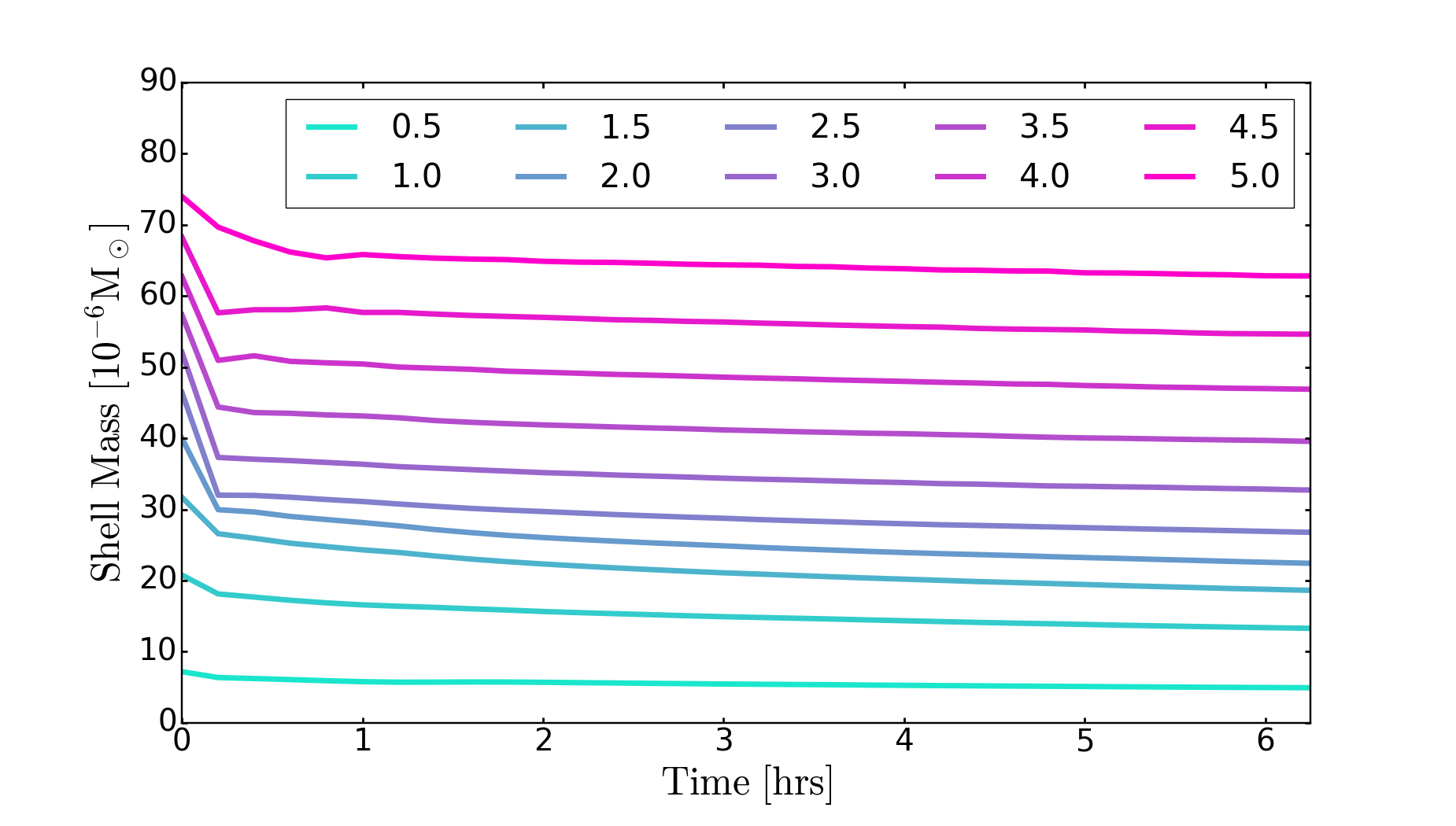}
    \caption{disc mass inside a cylindrical shell of thickness $\mathrm{0.5\times10^9 cm}$ for the 3 Jupiter mass disc. The colours indicate the outer radius in units of $\mathrm{10^9 cm}$. The values are independent of the cylindrical shell height of the cylinder as long as it is greater than the disc height\label{fig:shell}}
\end{figure}

Each disc is initially rotating at the Keplerian speed with the ambient stellar interior stationary. As such, the first few orbits are dominated by shear at the interface of the disc edges with the AGB interior. Because the disc is not in perfect HSE with the ambient at the start of the computation, the disc experiences some compression during the first quarter-orbit as seen in Figure~\ref{fig:shell} after which it reaches a quasi-steady state.  As the point particle cannot accrete, the central pressure increases, effectively halting accretion.  The result is mass outflow from the disc radially.

\begin{figure}
    \centering
    \includegraphics[scale=0.17]{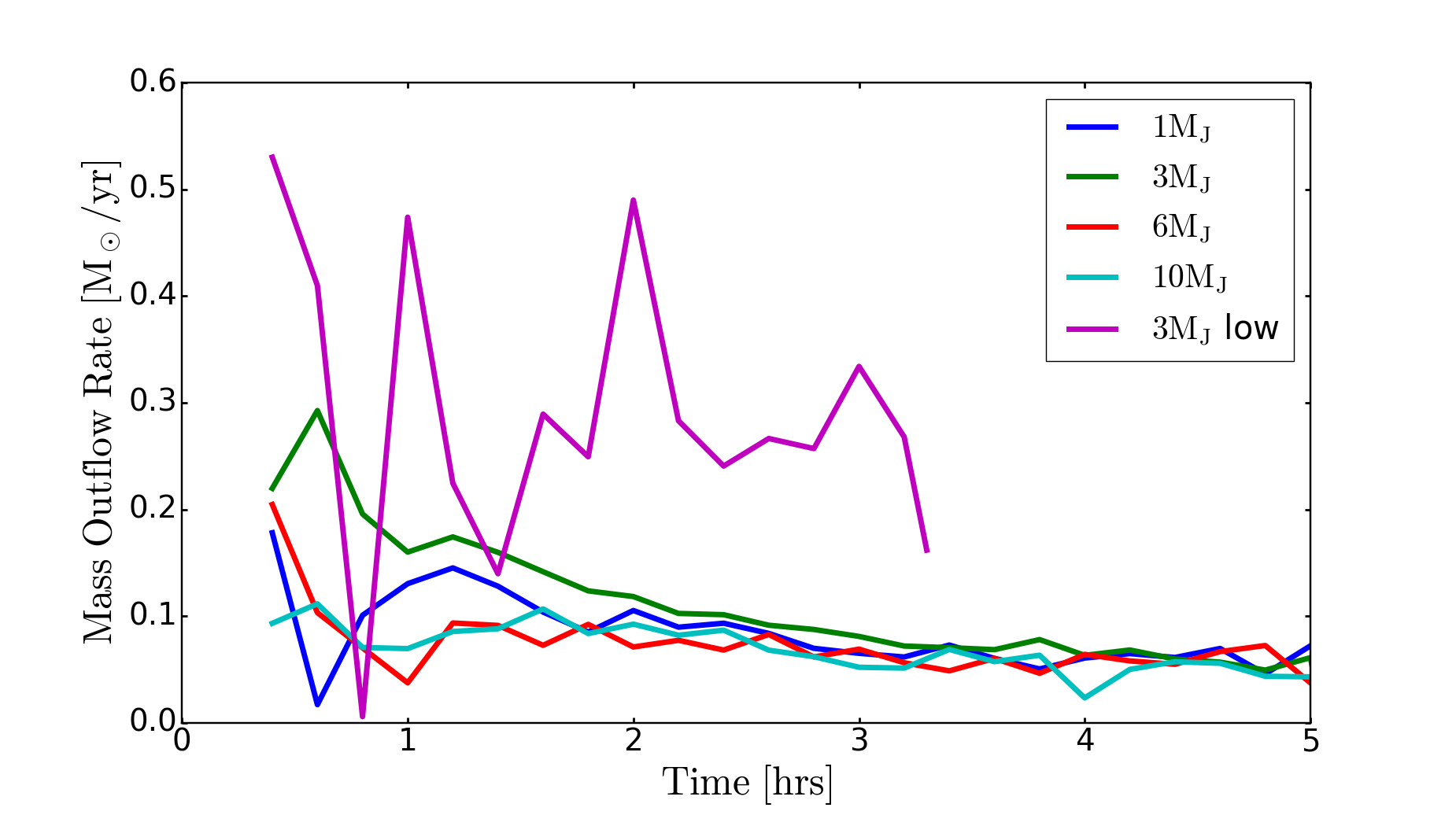}
    \caption{disc mass outflow rate inside a cylinder of radius $5\times 10^9$ cm centered on the AGB core after the disc reaches a steady state. The violet line is a lower resolution version of the $3$ Jupiter mass disc and shows that with the increased resolution we experience slower and more stable mass outflow.
    \label{fig:accrete}}
\end{figure}

Figure \ref{fig:shell} shows the disc mass inside of hollow cylinders with outer radius given by the colour key in units of $10^9$ cm, and with a thickness of $5\times 10^8$cm. In Figure~\ref{fig:accrete}, we present the mass outflow rate of the disc, i.e. the rate of change of the total disc mass inside a cylinder of radius $5\times 10^9$cm.  In aggregate, these simulations show that the discs become stable after a brief relaxation phase even in an environment where the initial shear is maximized.  Note that global 3D simulations of CE phases (the precursor of our systems) demonstrate that as the companion inspirals, the ambient stellar gas spins up via angular momentum transfer and can approach co-rotation with the disc \citep{Ricker_2008,Ricker2012,Iaconi_2018}. The faster the rotation of the surrounding AGB star, the less shear there would be between the disc and ambient material, and therefore more likely that the discs survive on long timescales.  Thus, we conclude that the stability of the discs formed is not threatened by the entrainment of hot material in a strongly sheared, realistic AGB environment. 

Our selected resolution for the simulations was the highest allowable given our computational budget and significantly higher than the highest resolution seen in global CE simulations \citep{Ohlmann2015,Chamandy:2018aa,Chamandy_2019,Passy2012}.  We check the robustness of our results by lowering the resolution of the 3 $M_{\rm J}$ simulation by two levels of refinement, i.e.~a factor of 4.  The mass outflow rate for the low resolution simulation is shown in Figure \ref{fig:accrete}.  Because the mass outflow rate is larger with lower resolution, it is likely that increasing our resolution would further decrease the mass outflow rate and thus increase the disc survival timescales. Note that the lower resolution simulation is far less steady compared to the higher resolution simulations which also suggests that the discs may further stabilize at higher resolution.
In summary, it is apparent that the discs persist for at least 10 orbits and perhaps $\gtrsim$100 orbits in all cases if the outflow rates remain constant.  Thus any magnetic field generated in the disc would likely have sufficient time to reach the surface of the white dwarf if the central pressure can be relieved via an outflow as was suggested in \cite{Nordhaus:2011aa}.


\section{Conclusions}
In this work, we present 3D adaptive-mesh-refinement hydrodynamic simulations of accretion discs around the core of an AGB star. These discs are expected to form from the tidal disruption of a low-mass companion inside a post-main-sequence star during a common envelope interaction. Such discs are initially cold and dense compared to the hot stellar ambient meaning entrainment of hot gas could dissolve the discs. Our simulations show that despite significant shear and temperature gradients, planetary mass discs could survive and operate on timescales long enough to amplify strong magnetic fields.  

Transport of the magnetic fields to the white dwarf surface requires a valve that can relieve central pressure.  As opposed to neutron stars and black holes who can remove pressure via neutrino cooling and advection through an event horizon, white dwarfs can in principle relieve pressure through strong outflows or jets.  Developing a sub-grid point-particle model that conserves thermal energy and appropriately couples to outflows would allow one to investigate whether the strong magnetic fields generated in the disc can anchor to the proto-white dwarf.

Future studies could also improve upon this work in several aspects.  Instead of starting with a well-formed disc, simulations that follow the inspiral of a self-gravitating planet as it tidally disrupts and settles into a disc would improve estimates of the initial disc mass and structure.  Furthermore, utilizing the full-MHD capability of {\sc AstroBEAR} in a tidal disruption simulation, or in the simulation setup described in this work, would allow one to study the amplification and dynamics of the magnetic field.

\section*{Acknowledgements}
GG and JN acknowledge support from the following grants: NASA HST-AR-15044, NASA HST-AR-14563, NTID SPDI-15992.  This work used the Extreme Science and Engineering Discovery Environment (XSEDE), which is supported by National Science Foundation grant number ACI-1548562. The Center for Integrated Research Computing (CIRC) at the University of Rochester provided additional computational and visualization resources. 
 
{\rm VisIt} is supported by the Department of Energy with funding from the Advanced Simulation and Computing Program and the Scientific Discovery through Advanced Computing Program.




\bibliographystyle{mnras}
\bibliography{MNRASpaper} 

\bsp	
\label{lastpage}
\end{document}